\documentstyle[12pt]{article}
\begin{document}
\begin{center}
{\large \bf LORENTZ SYMMETRY VIOLATION,}\\[0pt] \vspace{0.1cm}
{\large \bf VACUUM AND SUPERLUMINAL PARTICLES}\\[0pt]
\vspace{0.5cm} {\bf Luis GONZALEZ-MESTRES}\footnote{%
E-mail: lgonzalz@vxcern.cern.ch}
\\[0pt]
\vspace{0.3cm}
{\it Laboratoire de Physique Corpusculaire, Coll\`ege de France \\
11 pl.
Marcellin-Berthelot, 75231 Paris Cedex 05, France
\\[0pt] and
\\[0pt]
Laboratoire d'Annecy-le-Vieux de Physique des Particules \\ B.P. 110 , 74941
Annecy-le-Vieux Cedex,
France} \vspace{0.8cm}
\end{center}

\begin{abstract}
If textbook Lorentz invariance is actually
a property of the equations describing a sector
of the excitations of vacuum above some critical distance scale,
several sectors of matter with different
critical speeds in vacuum can coexist and an absolute rest frame (the vacuum
rest frame)
may exist without contradicting the apparent Lorentz invariance felt by
"ordinary" particles (particles with critical speed in vacuum equal to $c$ ,
the speed of light). Sectorial Lorentz invariance, reflected by the fact that
all particles of a given dynamical sector have the same critical speed in
vacuum, will then be an expression of a fundamental sectorial symmetry
(e.g. preonic grand unification or extended supersymmetry) protecting a
parameter of the equations of motion. Furthermore, the sectorial Lorentz
symmetry may be only a low-energy limit, in the same way as the relation
$\omega $ (frequency) = $c_s$ (speed of sound) $k$ (wave vector) holds for
low-energy phonons in a crystal. We show that, in this context, phenomena
such as the absence of Greisen-Zatsepin-Kuzmin cutoff and the stability
of unstable particles at very high energy are basic properties of a wide
class of noncausal models where local Lorentz invariance is broken
introducing a fundamental length. Then, observable phenomena are produced
at the wavelength scale of the highest-energy cosmic rays or even below this
energy, but Lorentz symmetry violation remains invisible to standard low-energy
tests. We discuss possible theoretical, phenomenological, experimental and
cosmological
implications of this new approach to matter and space-time, as well as 
prospects for future developments. 
\end{abstract}


\section{Introduction}

{\it ~~~"The impossibility to disclose experimentally the absolute motion
of the earth seems to be a general law of Nature"
\vskip 1mm
H. Poincar\'e
\vskip 3mm
"Precisely Poincar\'e proposed investigating what could be done with the
equations without altering their form. It was precisely his idea to pay
attention to the symmetry properties of the laws of Physics"
\vskip 1mm
R.P. Feynman
\vskip 3mm
"The interpretation of geometry advocated here cannot be directly applied
to submolecular spaces... it might turn out that such an extrapolation is
just as incorrect as an extension of the concept of temperature to particles
of a solid of molecular dimensions"
\vskip 1mm
A. Einstein}
\vskip 3mm
Is relativity the result of a symmetry of the laws of Nature (Poincar\'e,
1905), therefore necessarily broken at some deeper level (Einstein, late
period), or does it reflect the existence of an absolute space-time
geometry that matter cannot escape (Einstein, early papers on relativity)?
Most textbooks teach "absolute" relativity (early Einstein papers) and
ignore the possibility of a more flexible formulation (Poincar\'e, late
Einstein thought) that we may call "relative" relativity (relativity is a
symmetry of the laws of Nature expressed by the Lorentz group: whether this
symmetry is exact or approximate must be checked experimentally at each new
energy scale). 
In the first case, ether does not exist: light just propagates at the
maximum speed allowed by the "absolute" space-time geometry; in the
second case, the question of ether remains to be settled experimentally at 
any new small-distance scale. By introducing important dynamics into
the vacuum structure, particle physics has operated a return to ether: 
it would be impossible for the $W^{\pm }$ and the $Z^0$ to be gauge bosons 
with nonzero masses if they did not propagate in a medium where the Higgs
fields condense; similarly, modern theories of hadron structure conjecture 
that free quarks can exist only inside hadrons, due to non-trivial properties
(e.g. superconducting) of the non-perturbative QCD vacuum.
It could still be argued that this new "ether" does not 
necessarily have a preferred rest frame, and that special relativity is
an exact symmetry which prevents us from identifying such a frame.
However, this hypothesis does not seem to naturally fit which general
physics considerations. 
Modern dynamical systems provide many examples where Lorentz
symmetry (with a critical speed given by the properties of the system) is
a scale-dependent property which fails at the fundamental distance scale
of the system (e.g. a lattice spacing). In practical examples, the critical
speed of the apparently relativistic dynamical system is often less than
$10^{-5}~c$ and "relativity", as felt by the dynamical system, would forbid
particle propagation at the speed of light. Light would appear to such a
system just like superluminal matter would appear to us. 

Furthermore, high-energy physics has definitely found cosmic-ray events
with energies above $10^{20}~eV$ (e.g. Hayashida et al., 1997). 
This energy scale 
is, in orders of magnitude, 
closer to Planck scale ($10^{28}~eV$) than to the electroweak scale 
($10^{11}~eV$). Therefore, if Lorentz symmetry is not an exact symmetry of 
nature and is instead broken at $\approx ~10^{-33}~cm$ length scale, 
the parameters of Lorentz symmetry violation observed (if ever) 
in the analysis of the highest-energy cosmic-ray events will provide 
us with direct and unique information on physics 
at Planck scale. This may be the most fundamental physics outcome of
experiments such as AUGER (AUGER Collaboration, 1997) 
devoted to the study of cosmic rays at $E~
\approx ~10^{20}~eV$~: 
Lorentz symmetry violation at 
these energies would unravel phenomena originating at higher 
energy scales, including the possible existence of a fundamental length scale.
In the vacuum rest frame, particles of the same type moving at different
speeds are different physical objects whose properties cannot be made identical 
through a Lorentz transformation. This essential property remains true in
any other frame, but parameters measured in the vacuum rest frame have an
absolute physical meaning.
Indeed, assuming that the laboratory frame moves slowly with respect to 
the vacuum rest frame (which may be close to that suggested by the study of
cosmic microwave backgroud radiation),
the observed 
properties of particles at $E~\approx ~10^{20}~eV$ may look closer 
to physics at Planck scale than to physics at electroweak or GeV scale
(e.g. the failure of the parton model and of standard 
relativistic formulae for Lorentz contraction and time dilation; 
see Gonzalez-Mestres, 1997h) , and basic parameters of 
Planck-scale physics may become measurable through $E~\approx ~10^{20}~eV$
cosmic-ray events if Lorentz symmetry is violated (with exact Lorentz symmetry,
collisions of very high-energy cosmic rays would on the contrary be 
exactly equivalent
to collider events at much lower laboratory energies).

We review and comment
here recent work by the author on Lorentz symmetry violation
and possible superluminal sectors of matter.
Non-tachyonic superluminal particles (superbradyons) 
have been discussed in previous papers 
(Gonzalez-Mestres, 1995 , 1996 , 1997a and 1997b) and other papers have been
devoted to Lorentz symmetry violation (Gonzalez-Mestres, 1997c, 1997d and
1997e) as well as to its astrophysical consequences (Gonzalez-Mestres,
1997f and 1997g), its application to extended objects
(Gonzalez-Mestres, 1997h) and its relevance for future accelerator programs
(Gonzalez-Mestres, 1997i).

\section{Lorentz symmetry as a low-energy limit} 

Lamoreaux, Jacobs, Heckel, Raab and Forston (1986) 
have set an experimental limit from nuclear magnetic resonance 
measurements
which, when suitably analyzed (Gabriel and Haugan, 1990), 
amounts to the bound $\mid c _{matter}~-~c_{light}\mid ~<~6.10^{-21}
~c_{light}$ in the
$TH\epsilon \mu $ model of Lorentz symmetry violation (e.g. Will, 1993). 
However, the $TH\epsilon \mu $ model assumes a scale-independent
violation of Lorentz invariance (through the non-universality of the
critical speed parameter) which does not naturally emerge from
dynamics violating Lorentz symmetry at Planck scale or at some other  
fundamental length scale, where we would naturally expect such an effect 
to be scale-dependent and possibly vary (like the 
effective gravitational coupling)
according to a $E^2$ law ($E$ = energy scale). The $E^2$ law is indeed a 
trivial and rather general consequence of phenomena such as nonlocality,
as can be seen from the 
generalized one-dimensional Bravais lattice equation
(Gonzalez-Mestres, 1997d): 
\equation
d^2/dt^2~~[\phi ~(n)]~~~=~~~-~K~[2~\phi ~(n)~-~\phi ~(n-1)~-~\phi ~(n+1)]~~-~~
\omega _{rest}^2~\phi
\endequation
where $n$ (integer) stands for the site under consideration, 
$\phi (n)$ is a complex order parameter, $K$ an elastic constant
and $(2\pi )^{-1}~\omega _{rest}$ the frequency of the 
chain of oscillators in the zero-momentum limit. In the limit where the
lattice spacing vanishes but $K~a^2$ remains 
fixed, (1) becomes a two-dimensional dalembertian equation
of the Klein-Gordon type, with two-dimensional Lorentz symmetry and
critical speed parameter $K^{1/2}~a$ . 
In terms of the wave vector $k$ and the frequency $(2\pi )^{-1}~\omega $ ,
equation (1) leads to the dispersion relation:
\equation
\omega ^2~~(k)~~=~~2~K~[1~-~cos~(k~a)]~+~\omega _{rest}^2~~=
~~4~K~sin^2~(ka/2)~+~\omega _{rest}^2
\endequation
equivalent to:
\vskip 2mm
$E^2~~=~~2~K~[1~-~cos~(2\pi ~h^{-1}~p~a)~+~(2\pi )^{-1}~h
~\omega _{rest}^2]~~=$
\equation ~~~~~~~~~~~~~~~~~~~~~~~~~~~~~~~~~=~~4~K~[sin^2~(2\pi ~h^{-1}~p~a)~
+~(2\pi )^{-1}~h~\omega _{rest}^2]
\endequation
where $p$ stands for momentum and $h$ is the Planck constant.
The same procedure can be extended to a wide class of nonlocal models, 
inclundig those with continuous space, giving:
\equation
E~~=~~(2\pi )^{-1}~h~c~a^{-1}~e~(k~a)
\endequation
where $[e~(k~a)]^2$ is a convex
function of $(k~a)^2$ obtained from vacuum dynamics.
We have checked that this is also a fundamental property of old
scenarios breaking local Lorentz invariance (f.i. R\'edei, 1967),
although such a
phenomenon seems not to have been noticed by the authors.
Expanding equation (4) for $k~a~\ll ~1$ , we can write:
\equation
e~(k~a)~~\simeq ~~[(k~a)^2~-~\alpha ~(k~a)^4~+~(2\pi ~a)^2~h^{-2}~m^2~c^2]^{1/2}
\endequation
where $\alpha $ is a model-dependent constant, in the range $0.1~-~0.01$ for
full-strength violation of Lorentz symmetry at the fundamental length scale
($\alpha ~=~1/12$ for the Bravais-lattice model and its isotropic extension
to three dimensions),
and
\equation
E~~\simeq ~~p~c~[1~+~\alpha ~(k~a)^2/2]~+~m^2~c^3~(2~p)^{-1}
\endequation
and the new term $\Delta ~E~=~-~p~c~\alpha ~(k~a)^2/2$ in the right-hand 
side of (6) implies a Lorentz symmetry violation in the ratio $E~p^{-1}$
varying like $\Gamma ~(k)~\simeq ~\Gamma _0~k^2$ where $\Gamma _0~
~=~-~\alpha ~a^2/2$ . Such an expression is not incompatible with a possible
gravitational origin of Lorentz symmetry violation, where the effective
gravitational coupling would rise like $E^2$ below Planck energy. However,
other interpretations are possible (e.g. Gonzalez-Mestres, 1997d) where 
all presently kown "elementary" particles and gauge bosons would actually
be composite objects made of superluminal matter at Planck scale. 

More generally, a $k^2$ law for the parameters of Lorentz symmetry violation,
as suggested by the above formulae, would lead to substantial changes with
respect to conventional models (e.g. Will, 1993). In particular,  
Lorentz symmetry would remain unbroken at $k~=~0$ .
With such a law, an effect of order 1 at $p~=~3.10^{20}~eV~c^{-1}$ 
(the estimated 
momentum of the highest-energy observed cosmic-ray event) would become of order
$\approx ~10^{-25}$ at $p~=~100~MeV~c^{-1}$ (the highest momentum scale 
involved in nuclear magnetic resonance tests of special relativity). Therefore,
very large deviations from special relativity at the highest observed 
cosmic-ray energies would be compatible with a great accuracy of this theory
in the low-momentum region. Thus,  the main and most fundamental
physics outcome of very high-energy cosmic-ray
experiments involving particles and nuclei
may eventually be the test of special relativity. If Lorentz
symmetry is violated at Planck scale, the highest-energy cosmic ray events
may, if analyzed closely and with the expected high statistics from future
experiments, provide a detailed check of different models of deformed
relativistic kinematics and of the basic physics behind the kinematics. 

The same kind of deformed relativistic kinematics 
arises naturally in soliton models (Gonzalez-Mestres, 1997h). 
Starting from the equation:
\equation
c^{-2}~\partial ^2\psi /\partial t^2~~-~~\partial ^2\psi /\partial x^2~~~
=~~~2~\Delta ^{-2} ~\psi ~(1~-~\psi ^2)
\endequation
where $\Delta $ is the distance scale characterizing the soliton size
($x$ = space coordinate, $t$ = time coordinate), and
writing down the one-soliton solution of this equation:
\equation
\psi (x~,~t)~~=~~\Phi (y)~~=~~tanh~(\lambda _0~y)
\endequation
where $y~=~x~-v~t$ , $v$ is the speed of the soliton, $\lambda _0~=
~\Delta ^{-1}~\gamma _R$ and $\gamma _R$ 
is the standard relativistic Lorentz factor 
$\gamma _R~=~(1~-~v^2~c^{-2})^{-1/2}$ ,
we can introduce a perturbation to the system by adding to the 
left-hand side of
(7) a term $-~(a^2/12)~\partial ^4\psi /\partial~x^4$ 
which corresponds to the lowest-order correction to the
continuum limit when the Bravais-lattice version of (7) is expanded
in powers of $a^2$ . 
This new term in the equation will be compensated at the
first order in the perturbation by the replacement:
\equation
\Phi ~~\rightarrow ~~\Phi ~+~\epsilon ~\Phi~(1~-~\Phi ^2)
\endequation
where $\epsilon ~\propto ~a^2$ .
We furthermore replace $\lambda _0$ by a new coefficient $\lambda $ to
be determined form the perturbed equation. To first order in the perturbation, 
we get the solutions:
\equation
\epsilon~~\simeq ~~1~-~\lambda ^2~\gamma _R^{-2}~\Delta ^2
\endequation
\equation
\lambda ^2~~\simeq ~~[3~\pm ~(1~-~4~a^2~\Delta ^{-2}~\gamma _R^4/3)^{1/2}]~
(1~+~a^2~\Delta ^{-2}~\gamma _R^4/6)^{-1}~\Delta ^{-2}~\gamma _R^2/4
\endequation
leading for $\epsilon ~\ll ~1$
to $(\Delta ~\lambda )^{-2}~\simeq ~\gamma _R^{-2}~+
~a^2~\Delta ^{-2}~\gamma _R^2/3$ . Thus, 
relative corrections to standard relativistic
Lorentz contraction and time dilation factors are proportional to
$a^2~\Delta ^{-2}~\gamma _R^4$ and dominate when this variable becomes
$\approx ~1$ . Above this value of $\gamma _R$ , we expect departures from
special relativity to occur at leading level in many phenomena. Deformed
kinematics can be obtained at the lowest order in the perturbation.
A simplified calculation, valid for a wide class of soliton models,
could be as follows. At the first order in the perturbation with respect
to special relativity, we start from an effective lagrangian
for soliton kinematics:
\equation
L~~=~~-~m~c^2~\gamma _R^{-1}~(1~-~\rho ~\gamma _R^4)
\endequation
where $\rho $ is a constant proportional to $a^2~\Delta ^{-2}$ , according to
(10) and (11). From this lagrangian, we derive the expression for the
generalized momentum:
\equation
p~~=~~m~\gamma _R~v~(1~+~3~\rho ~\gamma _R^4)
\endequation
from which we can build the hamiltonian:
\equation
H~~=~~p~v~-~L~~=~~m~c^2~(\gamma _R~+~3~\rho ~v^2~c^{-2}~\gamma _R^5~-
~\rho ~\gamma _R^3)
\endequation
which
leads to a deformed relativistic kinematics defined by the relation:
\equation
E~-~p~c~~=~~m~c^2~\gamma _R^{-1}~(1~+~v~c^{-1})^{-1}~-~\rho ~~m~c^2~
[3~v~c^{-1}~(1~+~v~c^{-1})^{-1}~+~1]~\gamma _R^3
\endequation
and, when expressed in terms of $p$ at $v~\simeq ~c$ 
and for small values of $\rho ~\gamma _R^4$ , can be approximated by:
\equation
E ~-~p~c~~\simeq ~~m~c^2~(2~p)^{-1}~-~5~\rho ~p^3~(2~m^2~c)^{-1}
\endequation
where
the deformation term $5~\rho ~p^3~(2~m^2~c)^{-1}$
differs from that obtained from phonon mechanics in the Bravais lattice
only by a constant factor $\eta ~\propto ~2~h^2~(2\pi ~m~c~\Delta)^{-2}$ .

Looking at the low-speed
limit of (15), we find a renormalization of the critical speed
parameter $c$ ,
$\delta c$ , such that $\delta c~c^{-1}~
\approx ~(\Delta /a)^{-2}~\approx ~10^{-40}$ for hadrons if
$a~\approx ~10^{-33}~cm$ and $\Delta ~\approx ~10^{-13}~cm$ . This effect 
(the only one which survives at $k~=~0$) is
much smaller than the effects contemplated by other authors (e.g. Coleman and
Glashow, 1997 and references therein) and, even assuming that it would be 
different for different particles, it
cannot be excluded by existing data which can only rule out values of 
$\delta c~c^{-1}$ above $\approx ~10^{-20}$ .

\section {Deformed relativistic kinematics}

Assuming that Lorentz symmetry is violated at Planck scale or at some other
fundamental length scale, how does the new kinematics apply to different
particles, nuclei, atoms and larger objects? Other versions of deformed 
relativistic kinematics led in the past to controversies (Bacry, 1993 ; 
Fernandez, 1996)
which can be resolved (Gonzalez-Mestres, 1997h) if the value of $\alpha $
depends on the object under consideration.
In the presence of a fundamental (super)symmetry, it may be reasonable to
assume that $\alpha $ has the same value for leptons and gauge bosons. 
>From the above example with solitons, we conclude 
(Gonzalez-Mestres, 1997h) that the value of $\alpha $
for hadrons is naturally of the same order
as for "elementary" particles, although not necessarily identical.
It can also be different for different hadrons. A crucial question is how to
extend deformed relativistic kinematics to nuclei and larger objects. Two
different simplified approaches can be considered:

- {\it Model i)} . Due to the very large size of atoms, as
compared to nuclei, the transition from
nuclear to atomic scale appears as a reasonable point to stop considering
systems as "elementary" from the point of view of deformed relativity.
$\alpha $ would then have a universal value for nuclei and simpler objects,
but not for atoms and larger bodies.

- {\it Model ii)} . The example with $\Phi ^4$ solitons suggests that
hadrons can have values of $\alpha $ close to that of leptons and
gauge bosons, and the
transition may happen continuously at fermi scale,
when going from nucleons to nuclei. Then, the value of $\alpha $ would be
universal (or close to it) for leptons, gauge bosons and hadrons 
(solitons) but follow
a $m^{-2}$ law for nuclei (multi-soliton bound states) and heavier systems,
the nucleon mass
setting the scale.

Experimental tests should be performed and equivalent dynamical systems
should be studied.
However,  {\it Model i)} would lack a well-defined criterium to separate
systems to which the deformed relativity applies with the same value of
$\alpha $ as for leptons and gauge bosons from those to which this kinematics
cannot be applied, and to characterize the transition between the two regimes.
The above obtained $m^{-2}~\Delta ^{-2}$ dependence of
the coefficient of the deformation term for extended objects,
as described in {\it Model ii)}, seems to provide a continuous transition
from nucleons to heavier systems, naturally
filling this gap. On the other hand, a closer analysis reveals that there is
indeed a discontinuity between nuclei and atoms, as foreseen in
{\it Model i)} . As long as the deformation term
in electron kinematics can be neglected
as compared to the electron mass term, we can consider that most of the
momentum of an atom is carried by the nucleus and {\it Model ii)} may provide
a reasonable description of reality. But, when the electron mass term becomes
small as compared to the part of the energy it would carry in a parton
model of the atom, such a description becomes misleading. To have the same
speed as a nucleon, the electron must then carry nearly the same energy and
momentum.
We therefore propose (Gonzalez-Mestres, 1997h) 
a modified version of {\it Model ii)}
with $\Delta \approx~10^{-13}~cm$ from hadrons and nuclei where, for atoms
and larger neutral systems, the coefficient of the deformation term would be
corrected by a factor close to 1 at low momentum and
to 4/9 at high momentum
if the number of neutrons is equal to that of protons. This model is
obviously approximate and should be completed by a detailed dynamical
calculation that we shall not attempt here. It assumes that electrically
neutral bodies can reach very high energies per unit mass, which is
not obvious: spontaneous ionization may occur at speeds (in the
vacuum rest frame) for which
the deformation term in electron kinematics becomes larger
than its mass term.

Then, for bodies heavier than hadrons, the effective value of $\alpha $ would
decrease essentially like $m^{-2}$ . Applying a similar mass-dependence
to the $\kappa $ parameter of a different
deformed Poincar\'e algebra considered by
previous authors (Bacry, 1993 and references therein), i.e. $\kappa~ \propto ~
m$ for large bodies, yields the relation:
\equation
F~(M_0~,~E_0)~~=~~F~(M_1~,~E_1)~~+~~F~(M_2~,~E_2)
\endequation
with:
\equation
F~(m~,~E )~~=~~2~\kappa ~(m)~~sinh~[2^{-1}~\kappa ^{-1}~(m)~E ]
\endequation
where $M~=~M_1~+~M_2$ , $\kappa (m)$ is our above mass-dependent version
of the
$\kappa $ parameter of the deformed Poincar\'e algebra used by these authors,
and $E_0$ is the energy of a system with mass $M$ made of two non-interacting
subsystems of energies $E_1$ and $E_2$ and with masses $M_1$ and $M_2$ .
Defining mass as an additive parameter, the rest energy $E_{i,rest}$
(in the vacuum rest frame) of particle $i$ ($i~=~0~,~1~,~2$)
is given by the equation:
\equation
M_i~c^2~~=~~2~\kappa ~(M_i)~~sinh~[2^{-1}~\kappa ^{-1}~(M_i)~E_{i,rest}]
\endequation
and tends to $M_i~c^2$ as $~\kappa ~(M_i)~\rightarrow ~\infty$ .
Equations (17) and (18) lead to additive relations for the energy of macroscopic
objects if the proportionality rule $\kappa ~(m)~\propto ~m$
is applied. From our
previous discussion with a different deformation scheme, such a choice
seems to naturally agree with physical reality. Then,
contrary to previous claims (Bacry, 1993; Fernandez, 1996),
the rest energies of
large systems would be additive and no macroscopic effect on the total mass
of the Universe would be expected.

\section {Phenomenological implications}

As initially stressed, very high-energy cosmic rays can open a unique window to
Planck scale if Lorentz symmetry is violated. Contrary to standard
prejudice which would suggest that energy-dependent 
effects of Lorentz symmetry violation
at Planck scale can be detected only at energies close to this scale,  
it turns out that such effects are detectable at the highest observed
cosmic-ray energies. As discussed in Section 2 , we expect standard 
relativistic formulae for Lorentz contraction and time dilation for
a proton to fail at energies such that $a^2~\Delta ^{-2}~\gamma _R^4
~\approx ~1$ , i.e. $E~\approx 10^{19}~eV$ for $a~\approx ~10^{-33}~cm$ 
and $\Delta ~\approx ~10^{-13}~cm$ (Gonzalez-Mestres, 1997h). Similarly,
with the same figures and taking $\alpha ~\approx ~0.1$ ,
the proton mass term $m^2~c^3~(2~p)^{-1}$ in the expression for the proton
energy becomes smaller than the deformation term 
$\Delta ~E~=~-~p~c~\alpha ~(k~a)^2/2$ for $E$ above 
$\approx ~8.10^{18}~eV$ and, even if both terms are very small as compared
to the total energy, kinematical balances (which depend crucially on these
nonleading terms) are drastically modified (Gonzalez-Mestres, 1997d). 
The standard
parton picture of hadrons 
is equally disabled by the new kinematics at very high 
energy (Gonzalez-Mestres, 1997h), due to the impossibility for 
"almost-free" constituents carrying arbitrary fractions of 
the total energy and momentum 
to travel at the same speed.
Apart from the failure of the standard parton model for hadrons
at wave vectors above 
$\approx ~(8\pi ^2~\alpha ^{-1})^{1/4}~(m~c~h^{-1}~a^{-1})^{1/2}$
(i.e. at energies above $\approx ~10^{19}~eV$
if $a~\approx 10^{-33}~cm$ , $\approx ~10^{20}~eV$ for
$a~\approx 10^{-35}~cm$
and $\approx ~3.10^{17}~eV$ for
$a~\approx 10^{-30}~cm$), the following new effects at leading level would 
occur assuming a universal value of $\alpha $ for leptons, gauge bosons
and hadrons:

a) The Greisen-Zatsepin-Kuzmin (GZK) cutoff on very high-energy cosmic
nucleons (Greisen, 1966;
Zatsepin and Kuzmin, 1966) does no longer apply (Gonzalez-Mestres,
1997d and 1997f). Very high-energy cosmic rays originating
from most of the presently
observable Universe can reach the earth and generate the highest-energy
detected events. Indeed, fits to data below $E~=~10^{20}~eV$ using standard
relativistic kinematics (e.g. Dova, Epele and Hojvat, 1997) predict a sharp
fall of the event rate at this energy, in contradiction with data
(Bird et al., 1993 and 1996; Hayashida et al., 1994
and 1997; Yoshida et al., 1995)
which suggest that events above $10^{20}~eV$ are produced at a
significant rate. Lorentz symmetry violation from physics at Planck scale
provides a natural way out. The existence of the cutoff for cosmic nuclei will
then depend crucially on the details of deformed relativistic kinematics,
beyond the accuracy of the present discussion. 

b) Unstable particles with at least two massive particles in the final state
of all their decay channels
(neutron, $\Delta ^{++}$ , possibly muons, charged pions and $\tau $'s, 
perhaps some nuclei...)
become stable at very high energy
(Gonzalez-Mestres, 1997d and 1997f).
In any case, many 
unstable particles live longer than naively expected with exact
Lorentz invariance and, at high enough energy,
the effect becomes much stronger than previously estimated for nonlocal models
(Anchordoqui, Dova, G\'omez Dumm
and Lacentre, 1997)
ignoring the small violation of relativistic kinematics. Not only particles
previously discarded because of their lifetimes can be candidates for the
highest-energy cosmic-ray events, but very high-energy
cascade development can be modified
(for instance, if the $\pi ^0$ lives longer at energies above $\approx ~
10^{18}~eV$, thus favoring hadronic interactions and muon pairs
and producing less
electromagnetic showers).

c) The allowed final-state phase space of two-body collisions is 
modified at very high energy when, in the vacuum rest frame
where expressions (4) - (6) apply,
a very high-energy particle
collides with a low-energy target (Gonzalez-Mestres, 1997d). Energy
conservation reduces the final-state
phase space and can lead to a sharp fall of cross sections
starting at incoming-particle wave vectors well below the
inverse of the fundamental length, essentially above
$E~\approx ~(E_T~a^{-2}~h^2~c^2)^{1/3}$ where $E_T$ is the energy of the
target. For $a~\approx ~10^{-33}~cm$~, this scale corresponds to:
$\approx ~10^{22}~eV$ if the target is a rest proton; $\approx ~10^{21}~eV$ if
it is a rest electron; $\approx ~10^{20}~eV$ for a $\approx ~1~keV$ photon,
and $\approx ~10^{19}~eV$ if the target is a visible photon.
For a proton impinging on a $\approx ~10^{-3}~eV$ photon from cosmic microwave
background radiation, and taking $\alpha ~\approx ~1/12$ 
as in the Bravais-lattice model, 
we expect the fall of cross sections to occur above $E~\approx ~
5.10^{18}~eV$, the critical energy where the derivatives of the mass term
$m^2~c^3~p^{-1}/2$ and of the deformation term $\alpha ~p~(k~a)^2/2$
become equal in the expression relating the proton
energy $E$ to its momentum $p$ . With $a~\approx ~10^{-30}~cm$~, still allowed
by cosmic-ray data (Gonzalez-Mestres, 1997e), the critical energy scale
can be as low as
$E~\approx ~10^{17}~eV$ ; with $a~\approx ~10^{-35}~cm$ (still compatible with
data in the region of the predicted GZK cutoff), it would be at
$E~\approx ~5.10^{19}~eV$ .
Similar considerations lead to a fall of
radiation under external forces (e.g. synchrotron radiation) above this
energy scale. In the case of
a very high-energy $\gamma $ ray, taking $a~\approx ~10^{-33}~cm$ ,
the deformed relativistic
kinematics inhibits collisions with  $\approx ~10^{-3}~eV$ photons
from cosmic background radiation above $E~\approx ~10^{18}~eV$~,
with $\approx ~10^{-6}~eV$ photons above
$E~\approx ~10^{17}~eV$ and with $\approx ~10^{-9}~eV$ photons above
$E~\approx ~10^{16}~eV$~. Taking $a~\approx ~10^{-30}~cm$ would lower these
critical energies by a factor 100
according to the previous formulae,
whereas the choice $a~\approx 10^{-35}~cm$ would raise
them by a factor of 20 . 

d) In astrophysical processes,
the new kinematics may inhibit phenomena such as GZK-like cutoffs,
decays,
radiation emission under external forces
(similar to a collision with a very low-energy target), momentum loss
(which at very high energy does not imply deceleration) through collisions,
production of lower-energy secondaries, 
photodisintegration of some nuclei... {\bf potentially solving all
the basic
problems raised by the highest-energy cosmic rays}
(Gonzalez-Mestres, 1997e and 1997g). Due to the fall of
cross sections, energy losses become much weaker than expected with
relativistic kinematics and astrophysical particles can be pushed to much
higher energies (once energies above $10^{17}~eV$ have been reached through
conventional mechanisms, synchrotron radiation and collisions with ambient
radiation may start to be
inhibited by the new kinematics); similarly, astrophysical
particles will be able to propagate to
much longer astrophysical distances, and many more sources
(in practically all the presently observable Universe) can
produce very high-energy cosmic rays reaching the
earth; as particle lifetimes are
much longer, new possibilities arise for the nature of these cosmic rays.
Models of very high-energy
astrophysical
processes cannot ignore a possible Lorentz symmetry violation
at Planck scale, in which case observable effects are predicted for the
highest-energy detected particles.

e) If the new kinematics can explain the existence of $\approx 10^{20}~eV$
events, it also predicts that, above some higher energy
scale (around $\approx 10^{22}~eV$ for $a~\approx ~10^{-33}~cm$), the fall of
cross sections will prevent many cosmic rays 
(leptons, hadrons, gauge bosons) from depositing most of its
energy in the
atmosphere (Gonzalez-Mestres, 1997e).
Such extremely high-energy particles will produce atypical events of
apparently much lower energy. New analysis
of data and experimental designs are required to explore
this possibility. Again, the interaction properties of nuclei will depend on 
the details of deformed kinematics. 

Velocity reaches its maximum at
$k~\approx ~(8\pi ^2~\alpha ^{-1})^{1/4}~(m~c~h^{-1}~a^{-1})^{1/2}$ .
Observable effects of local Lorentz invariance breaking
arise, at leading level, well below the critical
wavelength scale $a^{-1}$ due to the fact that, contrary to previous models
(f.i. R\'edei, 1967), we directly apply non-locality to particle
propagators and not only
to the interaction hamiltonian. In contrast with
previous
patterns (f.i. Blokhintsev, 1966), $s-t-u$ kinematics ceases to make sense
and the motion of the global system with respect to the vacuum rest frame plays
a crucial role. The physics of elastic two-body scattering will depend
on five kinematical variables.
Noncausal dispersion relations (Blokhintsev and Kolerov,
1964) should be reconsidered, taking into account the departure from
relativistic kinematics. As previously stressed (Gonzalez-Mestres,
1997d) , this apparent
nonlocality may actually reflect the existence of superluminal sectors of
matter (Gonzalez-Mestres, 1996) where causality would hold at the
superluminal level (Gonzalez-Mestres, 1997a). Indeed, electromagnetism
appears as a nonlocal interaction in the Bravais model of phonon dynamics,
due to the fact that electromagnetic signals propagate much faster than
lattice vibrations.

Very
high-energy accelerator experiments
(especially with protons and nuclei) can play a crucial role
in the test of a possible Lorentz symmetry violation. To fit vith cosmic-ray
events,
they should be performed in the very-forward region. At
LHC, FELIX (e.g. Eggert, Jones and Taylor, 1997)
could provide a crucial check of special relativity by comparing
its data with cosmic-ray data in the $\approx ~10^{16}~-~10^{17}~eV$ region.
VLHC experiments would be expected to lead to fundamental studies in the
kinematical region which, according to special relativity, would be equivalent
to the collisions of $\approx ~10^{19}~eV$ cosmic protons. With a 700 $TeV$
per beam $p~-~p$ machine, it would be possible to compare the very-forward
region of collisions with those of cosmic protons at energies up to
$\approx 10^{21}~eV$~. Thus, it seems necessary that all very high-energy
collider programs allow for an experiment able to cover secondary particles
in the far-forward and far-backward regions. A model-independent way to
test Lorentz symmetry between collider and cosmic-ray data sould be carefully
elaborated, but the basic phenomena involved in the case of Lorentz symmetry
violation can be (Gonzalez-Mestres, 1997d and 1997h):

{\it i)} failure of the standard parton model (in any version, even
incorporating radiative corrections and phase transitions);

{\it ii)} failure of the relativistic formulae for Lorentz contraction
and time dilation;

{\it iii)} longer than predicted lifetimes for some of the produced particles
(e.g. the $\pi ^0$).

The role of high-precision data from accelerators would then be crucial to
establish the existence of such phenomena in the equivalent cosmic-ray
events. To reach the best possible performance, cosmic-ray experiments
should (if ever feasible) install in coincidence very large-surface detectors
(providing also the largest-volume target) with very large-volume underground
or undewater detectors and with balloon or satellite devices able to study
early cascade development.
It would then be possible to perform unique
tests of special relativity involving violations due to 
phenomena at some fundamental scale close to Planck scale, and even to
determine the basic parameters of Lorentz symmetry violation (e.g. of
deformed kinematics) and of physics at the fundamental length scale.

\section {Superluminal particles}

Lorentz invariance can be viewed as a symmetry of the motion
equations, in which case no reference to absolute
properties of space and time is required and the
properties of matter play the main role
(Gonzalez-Mestres, 1996). In a two-dimensional
galilean space-time,
the equation:
\equation
\alpha ~\partial ^2\phi /\partial t^2~-~\partial ^2\phi /\partial x^2 = F(\phi )
\endequation
with $\alpha$ = $1/c_o^2$ and $c_o$ = critical
speed, remains unchanged under "Lorentz" transformations leaving
invariant the squared
interval:
\equation
ds^2 = dx^2 - c_o^2 dt^2
\endequation
so that matter made with solutions of equation (20)
would feel a relativistic space-time even if the real space-time is actually
galilean and if an absolute rest frame exists in the
underlying dynamics beyond the wave equation.
A well-known example is provided by the solitons of the sine-Gordon equation,
obtained taking in (20):
\equation
F(\phi ) = - (\omega _0/c_o)^2~sin~\phi
\endequation
where $\omega _0$ is a characteristic frequency of the dynamical system.
A two-dimensional universe made of sine-Gordon solitons plunged
in a galilean world would behave like a two-dimensional minkowskian
world with the laws of special relativity.
Information on any absolute rest frame would be lost by the solitons, as
if the Poincar\'e relativity principle (Poincar\'e, 1905)
were indeed a law of Nature, even if
actually the
basic equation derives from a galilean world with an
absolute rest frame. The actual structure of space and time
can only be found by going beyond the wave equation
to deeper levels of resolution.
At this stage, a  crucial question arises
(Gonzalez-Mestres, 1995):
is $c$ (the speed of light) the only critical speed in vacuum, are
there particles with a critical speed different from that of light?
The question clearly makes sense, as in a
perfectly transparent crystal it is possible to identify
at least two critical speeds: the speed of light and
the speed of sound. It has been shown (Gonzalez-Mestres,
1995 and 1996) that superluminal
sectors of matter can be consistently
generated
replacing in the Klein-Gordon equation the
speed of light by a new critical speed $c_i$ $\gg $ $c$
(the subscript $i$ stands for the $i$-th superluminal sector). All
standard kinematical concepts and
formulas (Schweber, 1961) remain correct, leading to particles with
positive mass and energy which are not tachyons.
We call them {\bf superbradyons} as, according to standard
vocabulary (Recami, 1978), they are bradyons with superluminal critical speed in
vacuum. The rest energy
of a superluminal particle of mass $m$ and critical speed $c_i$
will be given by the generalized Einstein equation:
\equation
E_{rest}~~=~~m~c_i^2
\endequation
Energy and momentum conservation will in principle not be
spoiled by the existence of several critical speeds in vacuum:
conservation laws will as usual hold for phenomena leaving the vacuum
unchanged. Each superluminal sector will have its own Lorentz invariance
with $c_i$ defining the metric.
Interactions between two different
sectors will break both Lorentz invariances. Lorentz
invariance
for all sectors simultaneously will at best be explicit
(i.e. exhibiting the diagonal sectorial Lorentz metric) in a single
inertial frame ({\bf the vacuum rest frame}, i.e. the "absolute" rest
frame). 
If superluminal particles couple
weakly to ordinary matter, their effect on the ordinary sector will occur at
very high energy and very short distance (Gonzalez-Mestres, 1997c), far from
the domain of successful
conventional tests of Lorentz invariance
(Lamoreaux, Jacobs, Heckel, Raab and Forston,
1986 ;
Hills and Hall, 1990). In particular, superbradyons naturally escape the
constraints on the critical speed derived in some specific models
(Coleman and Glashow, 1997; Glashow, Halprin, Krastev, Leung and
Pantaleone, 1997).
High-energy experiments can therefore
open new windows in this field.
Finding some track of a superluminal sector (e.g. through
violations of Lorentz invariance in the ordinary sector) may
be a unique way to experimentally discover the vacuum rest frame.
Furthermore, superbradyons can be the fundamental matter from which
Planck-scale strings would actually be built.
Superluminal particles lead to consistent cosmological models
(Gonzalez-Mestres, 1997d), where they may well provide
most of the cosmic (dark) matter. Although recent criticism
to this suggestion has been emitted in a specific model
on the grounds of gravitation theory (Konstantinov, 1997),
the framework used is
crucially different
from the multi-graviton approach suggested in our papers, where
each dynamical sector would generate its own graviton.

Conventional tests of special relativity are performed using low-energy
phenomena. The highest momentum scale involved in nuclear magnetic
resonance tests of special relativity is related to the energy of virtual
photons exchanged, which does not exceede the electromagnetic energy scale
$E_{em}~\approx ~\alpha _{em}~r^{-1}~\approx ~1~MeV$ , where $\alpha _{em}$
is the electromagnetic constant and $r$ the distance scale between
two protons in a nucleus. However, the extrapolation between the 1 $MeV$
scale and the $1~-~100~TeV$ scale (energies to be covered by LHC and VLHC)
may involve a very large number, making compatible
low-energy results with the possible existence of superluminal particles above
$TeV$ scale.
Assume, for instance, that between $E~\approx ~1~MeV$ and $E~\approx ~100
~TeV$
the mixing between an "ordinary" particle (i.e. with critical speed
in vacuum equal to the speed of light $c$ in the relativistic limit)
of energy $E_0$ and
a superluminal particle with mass $m_i$ , critical speed $c_i~\gg ~c$  and
energy $E_i$ is described in the vacuum rest frame
by a non-diagonal term in the energy matrix of the
form (Gonzalez-Mestres, 1997c):
\equation
\epsilon ~\approx ~\epsilon _0~p~c_i~\rho~(p^2)
\endequation
where $p$ stands for momentum,
$\epsilon _0$ is a constant describing the strength of the mixing
and $\rho~(p^2)~=~p^2~(p^2~+~M^2~c^2)^{-1}$ accounts for a threshold effect
with $M~c^2~\approx ~100~TeV$ due to dynamics.
Then, the correction to the energy of
the "ordinary" particle will be $\approx ~\epsilon ^2~(E_0~-~E_i)^{-1}$
whereas the mixing angle will be $\approx ~\epsilon ~(E_0~-~E_i)^{-1}$ .
Taking the rest energy of the
superluminal particle to be $E_{i,rest}~=~m_i~c_i^2~\approx ~1~TeV$, we get
a mixing $\approx ~0.5~\epsilon _0$ at $p~c~=~100~TeV$ ,
$\approx ~10^{-2}~\epsilon _0$ at $p~c~=~10~TeV$ and
$\approx ~10^{-4}~\epsilon _0$
at $p~c~=~1~TeV$ . Such
figures would clearly justify the search for superbradyons at LHC and VLHC
($E~
\approx ~100~TeV$ per beam) machines
provided low-energy bounds do not force $~\epsilon _0$ to be too small.
With the above figures, at $p~c~=~1~MeV$ one would have a correction to
the photon energy less than
$\approx ~10^{-32}~\epsilon _0^2~p~c_i$ which, requiring
the correction to the photon energy not to be larger than $\approx 10^{-20}$ ,
would allow for large values of
$\epsilon _0$ if $c_i$ is less than $\approx ~10^{12}~c$ .
In any case, a wide range of values of $c_i$ and $\epsilon _0$ can be explored.
More stringent bounds may come from corrections to the quark propagator at
momenta $\approx ~100~MeV$. There, the correction to the quark energy would
be bounded only by $\approx ~10^{-24}~\epsilon _0^2~p~c_i$ and requiring it
to be less than $\approx 10^{-20}~p~c$ would be equivalent to $\epsilon _0~
<~0.1$ for $c_i~=~10^6~c$~. Obvioulsy, these estimates are rough and a detailed
calculation of nuclear parameters using the deformed relativistic kinematics
obtained from the mixing would be required. It must be noticed that the
situation is fundamentally different from that contemplated in the $TH\epsilon
\mu $ formalism and, in the present case, Lorentz invariance can remain
unbroken in the low-momentum limit, as the deformation of relativistic
kinematics for "ordinary" particles is momentum-dependent. Therefore, it may
be a safe policy to explore all possible values of $c_i$ and $\epsilon _0$ at
accelerators (including other possible parametrizations of $\epsilon $)
without trying to extrapolate bounds from nuclear magnetic
resonance experiments.

The production of one or two (stable or unstable) superluminal particles
in a high-energy accelerator experiment is potentially able to yield very
well-defined signatures through the shape of decay products
or "Cherenkov" radiation
in vacuum events (spontaneous emission of "ordinary" particles).
In the vacuum rest frame, a
relativistic superluminal particle would have energy $E~\simeq ~p~c_i$ ,
where $c_i~\gg ~c$ is the critical speed of the particle. When decaying into
"ordinary" particles with energies $E_{\alpha }~\simeq ~p_{\alpha }~c$
($\alpha ~=~1,...,N)$ for a $N$-particle decay product), the initial energy
and momentum must split in such a way that very large momenta $p_{\alpha }~
\gg ~p$ are produced (in order to recover the total energy with "ordinary"
particles) but compensate to give the total momentum $p$ .This requires the
shape of the event to be exceptionally isotropic, or with two jets back
to back, or yielding several jets with the required small total momentum.
Similar trends will arise in "Cherenkov-like" events, and remain observable
in the laboratory frame.
It must be noticed that, if the velocity of the laboratory with respect to the
vacuum rest frame is $\approx ~10^{-3}~c$ , the laboratory velocity of
superluminal particles as measured by detectors (if ever feasible) would
be $\approx ~10^3~c$ in most cases (Gonzalez-Mestres, 1997a).

The possibility that superluminal matter exists, and that it plays
nowadays an important role in our Universe,
should be kept in mind when addressing the two basic questions
raised by the analysis of any cosmic ray event:
a) the nature and properties of the cosmic ray primary; b)
the identification (nature and position) of the source of the cosmic ray.
If the primary
is a superluminal particle, it will escape conventional criteria
for particle identification
and most likely produce a specific signature
(e.g. in inelastic collisions) different from those of ordinary
primaries.
Like neutrino events, in the absence of
ionization (which will in any case be very weak) we
may expect the event to start anywhere inside the detector.
Unlike very high-energy neutrino events,
events created by superluminal primaries can
originate from a particle having crossed the earth.
As in accelerator experiments (see the above discussion), 
an incoming, relativistic superluminal particle with momentum $p$ and
energy $E_{in} \simeq p~ c_i $ in the vacuum
rest frame, hitting an ordinary particle at rest,
can release most of its energy into
two ordinary particles or jets with
momenta (in the vacuum rest frame)
close to $p_{max}~=~1/2~p~c_i~c^{-1}$ and oriented back
to back in such a way
that the two momenta almost cancel, or into several jets with a very 
small total momentum, or into a more or less isotropic event with an
equally small total momentum.
Then, an energy $E_R \simeq E_{in} $
would be transferred to ordinary secondaries.
Corrections due to the earth motion
must be applied (Gonzalez-Mestres, 1997a) before defining the expected
event configuration in laboratory experiments (AUGER, AMANDA...).
At very high energy, such events
would be easy to identify in large-volume detectors, even at very small rate.
If the
source is superluminal, it can be located anywhere
(and even be a free particle) and will not necessarily be at the same place as
conventional sources
of ordinary cosmic rays. High-energy cosmic ray events
originating form superluminal sources
will provide hints on the location of such
sources and be possibly the only way to observe them.
At very high energies,
the GZK
cutoff 
does not in principle hold for
cosmic ray events originating from superluminal matter:
this is obvious if the primaries are superluminal particles
that we expect to interact very weakly with the cosmic microwave
background,
but is also true for ordinary primaries as we do not expect them to
be produced at the locations of ordinary sources and there is no upper
bound to their energy around $100~EeV$.
Besides "Cherenkov" deceleration, a superluminal cosmic
background radiation may exist
and generate its own GZK
cutoffs. However, if there are large amounts
of superluminal matter around us, they can be the main superluminal source
of cosmic rays reaching the earth.

\vskip 6mm
{\bf References}
\vskip 4mm
\noindent
Anchordoqui, L., Dova, M.T., G\'omez Dumm, D. and Lacentre, P.,
{\it Zeitschrift f\"{u}r Physik C} 73 , 465 (1997).\newline
AUGER Collaboration, "The Pierre Auger Observatory Design 
Report" (1997).\newline
Bacry, H. Marseille preprint CPT-93/P.2911 , available from KEK database.
\newline
Bird, D.J. et al., {\it Phys. Rev. Lett.} 71 , 3401 (1993).\newline
Bird, D.J. et al., {\it Ap. J.} 424 , 491 (1994).\newline
Blokhintsev, D.I. and Kolerov, G.I., {\it Nuovo Cimento}
34 , 163 (1964).\newline
Blokhintsev, D.I., {\it Sov. Phys. Usp.} 9 , 405 (1966).\newline
Coleman, S. and Glashow, S.L. , {\it Phys. Lett. B} 405 , 249 (1997).\newline
Dova, M.T., Epele, L.N. and Hojvat, C., Proceedings of the 25$^{th}$ 
International Cosmic Ray Conference (ICRC97), Durban, South Africa, Vol. 7 ,
p. 381 (1997).\newline
Fernandez, J.,  {\it Phys. Lett. B} 368 , 53 (1996).\newline
Eggert, K., Jones, L.W. and Taylor, C.C., Proceedings of ICRC97 , Vol. 6 ,
p. 25 (1997).\newline
Gabriel, M.D. and Haugan, M.P. , {\it Phys. Rev. D} 41 , 2943 (1990).\newline
Glashow, S.L., Halprin, A., Krastev, P.I., Leung, C.N. and Pantaleone, J.,
"Comments on Neutrino Tests of Special Relativity", 
{\it Phys. Rev. D} 56 , 2433 (1997). 
\newline
Gonzalez-Mestres, L., "Physical and Cosmological Implications of a Possible
Class of Particles Able to Travel Faster than Light", contribution to the
28$^{th}$ International Conference on High-Energy Physics, Warsaw July 1996 .
Paper hep-ph/9610474 of LANL (Los Alamos) electronic archive (1996).\newline
Gonzalez-Mestres, L., "Space, Time and Superluminal Particles",
paper physics/9702026 of LANL electronic archive (1997a).\newline
Gonzalez-Mestres, L., "Superluminal Particles and High-Energy Cosmic Rays",
Proceedings of the $25^{th}$ International Cosmic Ray Conference (ICRC97),
Vol.6 , p.109 . Paper
physics/9705032 of LANL electronic archive (1997b).\newline 
Gonzalez-Mestres, L., "Lorentz Invariance and Superluminal Particles", 
March 1997 ,
paper physics/9703020 of LANL electronic archive (1997c).\newline
Gonzalez-Mestres, L., "Vacuum Structure, Lorentz Symmetry and Superluminal
Particles", paper physics/9704017 of LANL
electronic archive (1997d).\newline
Gonzalez-Mestres, L., "Lorentz Symmetry Violation and Very High-Energy
Cross Sections", paper physics/9706022 of LANL electronic archive (1997e).
\newline
Gonzalez-Mestres, L., "Absence of Greisen-Zatsepin-Kuzmin Cutoff
and Stability of Unstable Particles at Very High Energy,
as a Consequence of Lorentz Symmetry Violation", Proceedings of ICRC97 ,
Vol. 6 , p. 113 . Paper
physics/9705031 of LANL electronic archive (1997f).\newline
Gonzalez-Mestres, L., "Possible Effects of Lorentz Symmetry Violation on the
Interaction Properties of Very High-Energy Cosmic Rays", paper
physics/9706032 of LANL electronic archive (1997g).\newline
Gonzalez-Mestres, L., "High-Energy Nuclear Physics with Lorentz Symmetry
Violation", paper nucl-th/9708028 of LANL electronic archive (1997h).\newline
Gonzalez-Mestres, L., "Lorentz Symmetry Violation and Superluminal Particles at Future Colliders", paper physics/9708028 of LANL electronic archive (1997i).
\newline
Greisen, K., {\it Phys. Rev. Lett.} 16 , 748 (1966).\newline
Hills, D. and Hall, J.L., {\it Phys. Rev. Lett.} 64 , 1697 (1990).\newline
Hayashida, N. et al., {\it Phys. Rev. Lett.} 73 , 3491 (1994).\newline
Hayashida, N. et al., Proceedings of ICRC97 , Vol. 4 , p. 145 (1997).\newline
Konstantinov, M.Yu., "Comments on the Hypothesis about Possible Class
of Particles Able to Travel faster than Light: Some Geometrical Models",
paper physics/9705019 of LANL electronic archive (1997).
\newline
Lamoreaux, S.K., Jacobs, J.P., Heckel, B.R., Raab, F.J. and Forston, E.N.,
{\it Phys. Rev. Lett.} 57 , 3125 (1986).\newline
Poincar\'e, H., Speech at the St. Louis International Exposition of 1904 ,
{\it The Monist 15} , 1 (1905).\newline
Recami, E., in "Tachyons, Monopoles and Related Topics", Ed. E. Recami,
North-Holland, Amsterdam (1978).\newline
R\'edei, L.B., {\it Phys. Rev.} 162 , 1299 (1967). \newline
Schweber, S.S., "An Introduction to Relativistic Quantum Field Theory",
Row, Peterson and Co., Evanston and Elmsford, USA (1961).
\newline
Will, C.M. "Theory and Experiment in Gravitational Physics", Cambridge 
University Press (1993).\newline
Yoshida, S. et al., Proc. of the $24^{th}$ International Cosmic Ray
Conference, Rome, Italy, Vol. 1 , p. 793 (1995).\newline
Zatsepin, G.T. and
Kuzmin, V.A., {\it Pisma Zh. Eksp. Teor. Fiz.} 4 , 114 (1966).
\end{document}